\documentclass[9pt,twocolumn,twoside]{opticajnl}
\journal{opticajournal} 

\setboolean{shortarticle}{true}

\usepackage{lineno}

\title{Doppler-free spectroscopy of the Cs $6\text{S}_{1/2}-7\text{P}_{3/2}$ atomic transition at
456 nm in a nanometric-thick vapor layer}

\author[1]{Armen Sargsyan}
\author[2,*]{Emmanuel Klinger}
\author[2]{Rodolphe Boudot}
\author[1]{David Sarkisyan}

\affil[1]{Institute for Physical Research – National Academy of Sciences of Armenia, 0204 Ashtarak-2, Armenia}
\affil[2]{Universit\'e Marie et Louis Pasteur, SUPMICROTECH, CNRS, Institut FEMTO-ST, F-25000 Besan\c con, France}

\affil[*]{Corresponding author: emmanuel.klinger@femto-st.fr}

\begin{abstract}
The features of Doppler-free resonances detected by probing the $^{133}$Cs atom $6S_{1/2}-7P_{3/2}$ transition at 456\;nm in a nanometric-thick vapor layer are investigated. The matrix element of this transition is about 11 times smaller than that of the Cs D$_2$ line (852\;nm).
When the vapor layer thickness is $\ell = \lambda/2 \simeq 230$\;nm, we observe Dicke narrowing of the lines, accompanied by a red frequency shift of the atomic transitions, which is attributed to atom-surface interactions.
Realizing optical pumping with $\ell\simeq 460$\;nm in a single-pass configuration, we observe Doppler-free resonances with a linewidth $<20\;$MHz,
located at the atomic transitions frequencies
with a correspondence of the amplitudes to the transition intensities. These narrow resonances are of interest for high-resolution spectroscopy and instrumentation, and could serve as a frequency reference.
\end{abstract}

\setboolean{displaycopyright}{true}

\begin{document}
\twocolumn
\maketitle

Many contemporary optics and atomic physics experiments are realized using alkali vapors such as Rb and Cs. 
Spectroscopy of hot vapors \cite{pizzey2022laser} is at the core of many devices and applications, such as atomic optical clocks and gyroscopes \cite{kitching2018chip}, optical magnetometers \cite{FabricantNJP2023}, Faraday filters \cite{Uhland2023}, or light-trapping experiments with electromagnetically induced transparency \cite{finkelstein2023practical}. Typically, the transitions used in these experiments are the D$_{1}$ and D$_{2}$ lines, resonant with near-infrared light (600 -- 900\;nm). 

Probing higher-frequency transitions in cesium with sub-Doppler spectroscopy has recently attracted interest as these transitions could serve as optical frequency references \cite{Miao2022,Klinger2024sub-doppler}. Nevertheless, the literature on high resolution spectroscopy of the $7\text{P}_{J}$ states of $^{133}$Cs is scarce. The Cs 6S$_{1/2}\rightarrow 7$P$_{J}$ transitions were investigated for both $J=1/2$ (459\;nm) and $J=3/2$ ($456\;$nm) to precisely measure their matrix element \cite{Damitz2019measurements} and their lifetime \cite{Toh2019}. 
A study of nonlinear magneto-optical resonances observed in the fluorescence to the ground state from the 7P$_{3/2}$ state, populated by direct optical pumping with a laser radiation at 456\;nm, was carried out in Ref.\;\cite{Auzinsh2011cascade}. In Ref.\;\cite{Li2019continously}, the $6\text{S}_{1/2}\rightarrow 7\text{P}_{3/2}$ transitions were studied using the well-known saturated absorption (SA) technique, enabling sub-Doppler resolution of these transitions. 

\begin{figure}[htb]
    \centering
    \includegraphics[width=\linewidth]{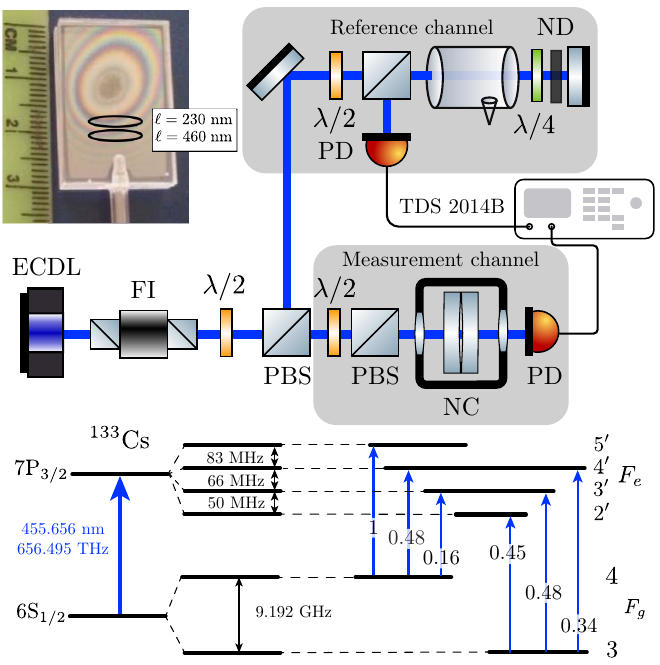}
    \caption{(top) Sketch of the experimental setup. ECDL -- external cavity diode laser; FI -- Faraday isolator; PBS -- Glan polarizer; NC -- nanocell containing Cs atomic vapor, placed inside the heater; PD--photodetectors. The signals are proceeded by a digital storage oscilloscope. The frequency reference is realized with saturated absorption spectroscopy. The inset shows a photograph of the NC filled with Cs where the upper and lower ovals mark the thickness $\ell \sim 230\;$nm and $\ell \sim 460\;$nm. (bottom) Energy levels involved in the $6\text{S}_{1/2}\rightarrow 7\text{P}_{3/2}$ transitions. The relative frequency shifts are indicated between hyperfine levels, together with the transition intensities, normalized to the largest one: $F_g=4\rightarrow F_e=5$.}
    \label{fig:1}
\end{figure}

As was thoroughly studied on alkali D lines, see e.g. Refs.\;\cite{dutierJOSAB2003,dutier2003collapse,sarkisyan2004spectroscopy,sargsyan2008novel}, and even some molecular lines \cite{Hartmann2016,arellanoNC2024}, probing resonances in vapor layers having a thickness on the order of the excitation wavelength offers an alternative to usual sub-Doppler spectroscopy achieved with nonlinear processes. One advantage is that light-matter interaction occurs in the weak-probe regime, simplifying both the experimental setup and the interpretation of results.  In these cells, referred to as nanocells (NC), the narrowest transmission spectra are obtained at two preferred thicknesses: $\ell = \lambda/2$ and $\ell = \lambda$, where $\lambda$ is the wavelength of laser radiation that is resonant with the corresponding transition. In the first case ($\ell = \lambda/2$), both the transmission and fluorescence spectra exhibit a significant sub-Doppler narrowing (by factors of three and six, respectively) compared to centimeter-sized vapor cells \cite{dutier2003collapse,sarkisyan2004spectroscopy}. The second case ($\ell = \lambda$) is characterized by the possibility of forming narrow resonances by means of velocity-selective optical pumping (VSOP), located at atomic transitions, in the transmission spectrum \cite{sargsyan2008novel}. 
To allow for the study of atomic resonances as a function of the thickness $\ell$ of the vapor layer (transverse to the laser beam), our NCs 
are fabricated with a wedge gap between their windows, in the vertical direction \cite{keaveney2012cooperative}. In this way, the required thickness 
is achieved either via mechanical movement of NC or via spatial translation of the laser beam. 

In this work, we perform qualitative Doppler-free spectroscopy of the $^{133}$Cs $6\text{S}_{1/2} \rightarrow 7\text{P}_{3/2}$ transition at $\lambda=455.6\;$nm in a nanometric thick vapor layer, contained in a NC. One of the challenges is that the $6\text{S}_{1/2} \rightarrow 7\text{P}_{3/2}$ transition has a dipole moment about 11 times smaller than that of the D$_2$ line \cite{Damitz2019measurements}, and, together with the small optical path of the NC, makes the recording of spectral features challenging. This issue is addressed by increasing the temperature of the NC to about 180$^\circ$C where it would usually be set to about 120$^\circ$C in the case of alkali D lines. The spectra obtained for $\ell=\lambda/2$ and $\ell=\lambda$ are compared with that obtained with saturated absorption spectroscopy from a cm-long vapor cell. When $\ell = \lambda/2$, we show that the Dicke narrowing is accompanied by a 35\;MHz-large red frequency shift and broadening of the atomic transitions due to atom-surface interactions, negligible at this scale in the case of D lines. This makes wedged NCs a platform of choice to perform quantitative studies of atom-surface interactions. We also show that the thickness $\ell=\lambda$ is preferred to perform high-resolution spectroscopy. In this case, the second derivative \cite{sargsyanOL2019} of the spectra exhibits sub 20\;MHz-wide Doppler-free resonances not affected by atom-surface interactions.

The experimental setup is sketched in Fig.\;\ref{fig:1}. A tunable external-cavity diode laser (ECDL) with a spectral linewidth of about 400\;kHz is tuned in the vicinity of the $6\text{S}_{1/2} \rightarrow 7\text{P}_{3/2}$ transitions ($\lambda=456\;$nm) of Cs atoms. The laser beam is directed at normal incidence onto the windows of a NC containing Cs atomic vapor. The NC is placed in an oven to adjust the temperature to about $T\approx 180^\circ$C, corresponding to a vapor number density of about $5.5 \times 10^{14}\;$cm$^{-3}$. The vapor column thickness is adjusted by translation of the nanocell-oven assembly. The thickness of atomic vapor column $\ell \approx\lambda/2$ or $\ell \approx\lambda$ are marked on the top left inset in Fig.\;\ref{fig:1}. The details of the NC design can be found in Ref.\;\cite{keaveney2012cooperative}. The light transmitted through the NC is recorded using photodiodes, and their signals are fed into a digital storage oscilloscope. A fraction of the light is used to make a frequency reference channel based on saturated absorption in a cm-long glass-blown Cs cell.

\begin{figure*}[htb]
    \centering
    \includegraphics[width=0.95\linewidth]{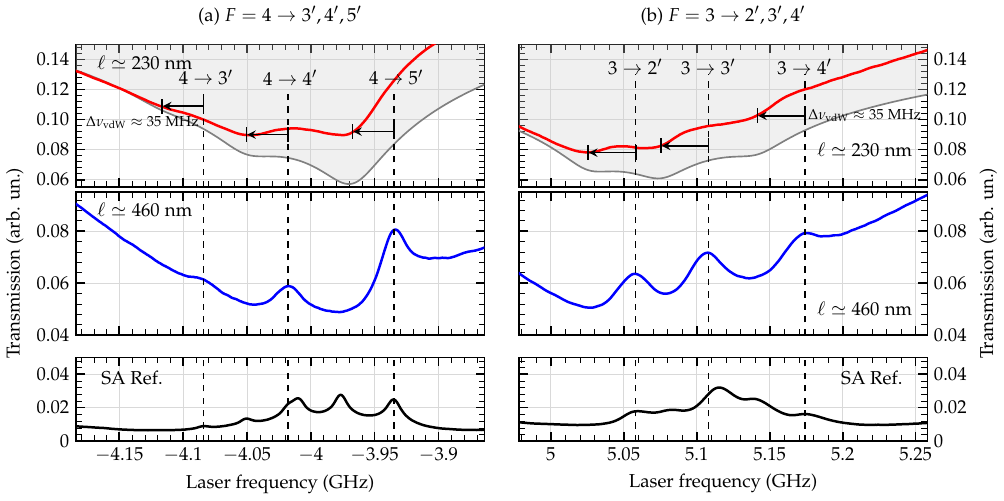}
    \caption{Transmission spectra of the $6\text{S}_{1/2} \rightarrow 7\text{P}_{3/2}$ transition at 455.6\;nm for (a) the $F=4\rightarrow 3',4',5'$ manifold and (b) the $F=3\rightarrow 2',3',4'$ manifold.
    The top panels show the transmission spectra obtained for a vapor layer thickness of $\ell\simeq 230\;$nm and a laser power of about 0.2\,mW. A 35\;MHz-large red frequency shift of the atomic transitions with respect to the unperturbed transitions marked by dashed lines is observed. This shift is attributed to atom-surface interaction \cite{Laliotis2021atom-surface}. The gray-filled curves are the theoretical spectra calculated for $\ell=230\;$nm with $\Gamma = 40\;$MHz, taking into account \eqref{eq:vdW}, see text. The middle panels show the transmission spectra obtained with $\ell\simeq 460$\;nm, which contains six VSOP resonances (three for each manifold) not shifted by atom-surface interaction, with a linewidth of about 28 MHz. These are obtained by increasing the laser power to about 1\;mW. The bottom panels show spectra obtained with saturated absorption from a cm-long glass-blown vapor cell. The six VSOP resonances, marked by vertical dashed lines, are seen to partially overlap with six CO resonances. The zero laser frequency corresponds to the weighted center of the $6\text{S}_{1/2} \rightarrow 7\text{P}_{3/2}$ transition. Note that the peak transmission in the case of the NC is 99\% while it is about 85\% in the case of the cm-long cell.}
    \label{fig:2}
\end{figure*}

Figure\;\ref{fig:2} shows the features of the transmission spectrum for thicknesses $\ell \approx \lambda/2$ and $\ell \approx \lambda$. Figure\;\ref{fig:2}(a) shows the transmission spectrum of the $F_g=4 \rightarrow F_e=3,4,5$ manifold (hereafter, the notation is simplified to $ F= 4 \rightarrow 3',4',5'$), while Fig.\;\ref{fig:2}(b) shows the transmission spectrum of the $F=3 \rightarrow 2',3',4'$ manifold. These transitions are indicated by vertical dashed lines. The energy level diagram, showing the transitions and their intensities, is presented at the bottom of Fig.\;\ref{fig:1}.


The top panels in Fig.\;\ref{fig:2} show the transmission spectra obtained with $\ell \approx \lambda/2$ and a laser power of about 0.2\;mW.
In this case, the spectra exhibit a significant narrowing, referred to as Dicke narrowing: the transmitted spectra are typically about four times narrower in comparison with the Doppler linewidth \cite{dutier2003collapse,sarkisyan2004spectroscopy}.
Here, the transitions are, in addition, seen to be shifted by about 35\;MHz toward lower frequencies with respect to the non-shifted atomic transitions. This is caused by van der Waals (vdW) interaction between Cs and the technical sapphire windows of the NC, evidence of atom-surface interaction \cite{Chevrollier1991vanderwaals,Laliotis2021atom-surface,Sargsyan2023competing,Dutta2024effects}. This is characterized by a broader, asymmetric atomic lineshape whose center experiences a red frequency shift. To estimate the red frequency shift in an NC, one can use the formula \cite{sargsyanOL2019}
\begin{equation}\label{eq:vdW}
   \Delta\nu_{\text{vdW}}=-16C_3 / \ell^3, 
\end{equation}
where $C_3$ is a coupling constant that depends on the surface and the atomic state. For alkali D lines interacting with sapphire, one typically has $C_3 \sim 1-2\;$kHz\;µm$^3$. For this reason, the 
VdW interaction causes a red frequency shift typically observable for $\ell<100\;$nm \cite{Peyrot2019measurement}. It is known, however, that the atomic polarizability, linked to the $C_3$ coefficient of vdW effect, is increasing with the principal quantum number $n$ \cite{Adams2019,Sargsyan2023competing}. Indeed, we measure $\Delta\nu_{\text{vdW}} \sim 35\;$MHz for the $6\text{S}_{1/2} \rightarrow 7\text{P}_{3/2}$ transition at $\ell \simeq 230\;$nm, yielding $C_3 \sim 20 \;$kHz\;µm$^3$. This is in good agreement with the value presented in Ref.\;\cite{Laliotis2021atom-surface}. 

The gray-filled curves in the top panels of Fig.\;\ref{fig:2} are theoretical spectra calculated following the approach described in Ref.\;\cite{dutierJOSAB2003} with $\Gamma /2\pi = 40\;$MHz and $\ell = 230\;$nm. Here, we have taken into account the van der Waals shift, given by \eqref{eq:vdW}. Note, however, that the red frequency shift due to AS interaction is accompanied by a distortion of the atomic resonance profile, which we do not take into account. This explains the discrepancies observed between experiments and theory.

Middle panels in Fig.\;\ref{fig:2} show transmission spectra obtained for $\ell\simeq 460\;$nm and a laser power of about 1\;mW. It contains six VSOP resonances (three for each ground state manifold) located at the corresponding transitions.
The VSOP resonance linewidth (FWHM) is measured to be 28\;MHz, which accounts for the natural linewidth $\Gamma_N/2\pi \approx 1.16\;$MHz of the 7P$_{3/2}$ level \cite{Toh2019} and some residual Doppler width.  To estimate the VSOP linewidth, one can use the simple formula 
\begin{equation}
    \Gamma_{VSOP}= \sqrt{\Gamma_{D} \times \Gamma_N} \approx 2\pi\; 32\;\text{MHz}.
\end{equation} 
Note that VSOP resonances can appear in the transmission spectrum obtained from a nanocell with thicknesses $\ell=m \lambda$, where $m$ is an integer. However, with increasing $m$, a spectral broadening occurs.

In comparison, the bottom panels in Fig.\;\ref{fig:2} show the saturation absorption spectrum obtained with a 1-cm long Cs vapor cell heated to about $50^\circ$C. For each ground state manifold, it contains three VSOP resonances and three crossover (CO) resonances, typical of saturated absorption. 
 Here, the transitions overlap because of the presence of CO resonances, often more intense than the transitions themselves. Meanwhile, the transitions are very well spectrally resolved in the case of the NC. Moreover, the amplitudes of the resonances correspond to the transition intensities, see Fig.\;\ref{fig:1}. Thus, VSOP resonances obtained from a NC with $\ell \approx \lambda$ appear promising to serve as a frequency reference tool.

\begin{figure*}[htb]
    \centering
    \includegraphics[width=0.95\linewidth]{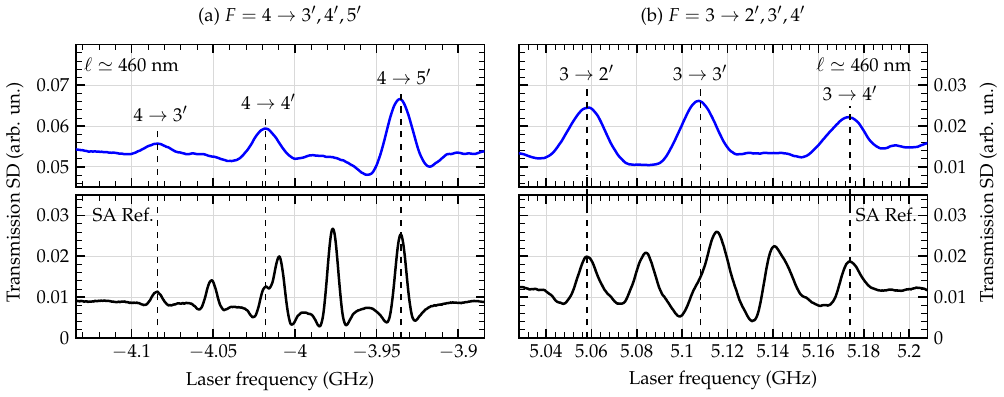}
    \caption{Second derivative transmission spectra of the $6\text{S}_{1/2} \rightarrow 7\text{P}_{3/2}$ transition at 455.6\;nm for (a) the $F=4\rightarrow 3',4',5'$ manifold and (b) the $F=3\rightarrow 2',3',4'$ manifold. The top panels show the SD spectra obtained after processing the raw transmission spectra recorded from a vapor layer of thickness $\ell\simeq 460\;$nm (middle panels in Fig.\;\ref{fig:2}). The bottom panels show the SD spectra obtained by processing the SA reference spectrum (bottom panels in Fig.\;\ref{fig:2}). All of these demonstrate a better spectral resolution than the raw transmission spectra.}
    \label{fig:3}
\end{figure*}

As was shown in \cite{sargsyanOL2019},  the second derivative (SD) technique can be employed to further narrow the atomic lines.  It can be done through either digital processing (for example, with a frequency modulation technique) of the recorded transmission trace or by numerical processing with a computer. Figure\;\ref{fig:3} shows the curves of the middle and bottom panels of Fig.\;\ref{fig:2} 
processed with second derivative.
In the case of the NC (top panels), the transition linewidth in the SD spectrum  is seen to reduce down to about 15\;MHz, that is a 58-fold narrower as compared with the Doppler width ($\Gamma_D/2\pi\approx 0.87\;$GHz at 180$^\circ$C). A similar value is obtained in Ref.\;\cite{Klinger2024sub-doppler} using SA technique with two counter-propagating laser beams in a 1.4-mm long microfabricated Cs cell. Here, these narrow resonances are obtained in a single-pass configuration. Note that the SD treatment of signals recorded with NCs preserves not only the frequency positions of the spectral features but also their amplitudes linked
to the transition probabilities. Thus, the narrow VSOP resonances obtained with the NC at $\ell \approx 460\;$nm are of interest for high-resolution spectroscopy and instrumentation.
The bottom panels in Fig.\;\ref{fig:3} show, for comparison, the SA transmission spectrum also processed with SD. Despite being narrower, spectral features corresponding to atomic transitions are still partially overlapped with CO resonances. That is especially the case for transitions $F=4\rightarrow4'$ [Fig.\;\ref{fig:3}(a)] and $F=3\rightarrow3'$ [Fig.\;\ref{fig:3}(b)].

The advantages of sub-Doppler spectroscopy using NCs as a frequency reference tool are as follows: $(i)$ simplicity of the realization geometry (single-beam transmission as opposed to counter-propagating beams requirement for SA geometry); $(ii)$ low value of the required laser power (typically a factor of ten less than that needed for the SA technique); $(iii)$ absence of CO resonances, which is especially important for example when atomic transitions are studied in an external field, since this leads to the splitting of the CO resonances into a large number of additional components; $(iv)$ correspondence of the amplitudes of atomic resonances to their corresponding transition intensities. 
Finally, it is important to note that achieving a precise NC thickness of $\ell = \lambda/2$ or $\ell = \lambda$ is not crucial: narrow spectra are observed within a tolerance of $\Delta\ell = \pm 30\;$nm, which makes the proposed technique experimentally feasible.

In conclusion, we have performed high-resolution spectroscopy of the 6S$_{1/2} \rightarrow 7$P$_{3/2}$ transitions at 455.6\;nm in a Cs vapor nanolayer. We have observed that the Dicke narrowing of the spectrum at $\ell = \lambda /2 \simeq 230\;$nm is accompanied by a red frequency shift due to atom-surface interactions. This makes wedged NCs a platform of choice to perform quantitative studies of atom-surface interactions.
In addition, the resonances formed using a NC with $\ell=\lambda \simeq 460\;$nm and VSOP, experience a 50-fold spectral narrowing with respect to the Doppler broadened Cs atomic lines. This last result could be useful to investigate the evolution of the transitions in a magnetic field. One expects to observe, for example, magnetically-induced transitions \cite{tonoyanEPL2018}: five transitions for the $F=4\rightarrow2'$ manifold and seven for the $F = 3\rightarrow5'$ manifold of the 6S$_{1/2} \rightarrow 7$P$_{3/2}$ transition.
We also expect that Cs NCs could be successfully used in the spectroscopy of 6S$_{1/2} \rightarrow 7$P$_{1/2}$ (459\;nm) 6S$_{1/2} \rightarrow 8$P$_{3/2}$ spectroscopy (388\;nm). Note that the recent development of a glass NC \cite{PeyrotOL2019}, which is easier to manufacture than sapphire-made NC used in the present work, could make the nanocells available for a wider range of researchers.

\section*{Funding}
AS and DS acknowledge support from the Science Committee of RA, in the frame of the research project of project n$^\circ$1-6/23-I/IPR. EK and RB acknowledge support from Agence Nationale de la Recherche (EIPHI Graduate school grant ANR-17-EURE- 0002); Région Bourgogne Franche-Comté.

\section*{Acknowledgments}

\section*{Disclosures}
The authors declare no conflicts of interest.

\section*{Data availability statement}
The data that support the findings of this study are available from the corresponding author upon reasonable request.

\bibliography{Bib-main}
\bibliographyfullrefs{Bib-main}

\end{document}